\documentstyle[12pt]{article}
\title{The effect of the neutron star crust on the evolution of a core magnetic 
        field}
\author{D.Konenkov$^1)$ and U.Geppert$^2)$  \\
     1) A.F.Ioffe Institute of Physics and Technology \\
        Politechnicheskaya, 26 \\
	194021 St.Petersburg, Russia \\
        {\it e-mail: dyk@astro.ioffe.rssi.ru} \\
     2) Astrophysikalisches Institut Potsdam \\
        An der Sternwarte 16 \\
        D-14482 Potsdam, Germany \\
        {\it e-mail: urme@aip.de} }

\begin{document}

\baselineskip15pt

\maketitle

\newpage
\begin{abstract}

We consider the expulsion of the magnetic field from the super--conducting
core of a neutron star and its subsequent decay in the crust.  Particular
attention is paid to a strong feedback of the distortion of magnetic
field lines in the crust on the expulsion of the flux from the core. This
causes a considerable delay of the core flux expulsion if the initial
field strength is larger than $10^{11}$ G. It is shown that the hypothesis
on the magnetic field expulsion induced by the neutron star
spin--down is adequate only for a relatively weak initial magnetic
field $B \approx 10^{11}$ G.  The expulsion time--scale depends not
only on the conductivity of the crust, but also on the
initial magnetic field strength itself. Our model of the field
evolution naturally explains the existence of the residual magnetic field
of neutron stars. Its strength is correlated with the impurity
concentration in neutron star crusts and anti--correlated with the
initial field strengths.

\vspace{4cm}

\end{abstract}

\newpage
\section{Introduction}

Where the neutron star magnetic field (MF) is located, by which processes  it
is generated and what determines its strength, structure and evolution, is the
subject of scientific debates since neutron stars are conceivable for  human
mind. Regarding the field location there exist basically two qualitatively 
different ideas: the field is either present in the entire star, that is, in the core
and the crust, or is located mainly in the crust. Whether the entire neutron star
or only the crust is penetrated by the field depends strongly on the  mechanism
responsible for its generation.

Both the simple idea of flux conservation during the gravitational collapse and
the action of a dynamo in the convective proto--neutron star (Thompson \& Duncan
1993) result in a field structure which penetrates the entire star.  On the
other hand it is possible that fall--back accretion after the supernova
submerges  any initially existent field (see Geppert,  Page \& Zannias
1999). Since the re--diffusion process may last a long time, the existence
of young pulsars (PSRs) could be explained  - at  least to some extent - by the
action of a thermoelectric instability (Blandford, Applegate \& Hernquist 1983, 
Urpin,  Levshakov \& Yakovlev 1986,
Wiebicke \& Geppert 1996).  A strong MF can be
produced rather rapidly in the surface layer of the neutron star by
transforming heat flux into magnetic flux. If so, the surface MF which governs e.g. the  spin--down of the PSR
is maintained by currents confined in the crust alone.  However, the currently
existing models describe only the amplification of large scale toroidal
fields, the production of the observed  dipolar MF is still an open problem.

Although the assumption of a purely crustal MF explains well the observed
long--term  evolution of isolated neutron stars (Urpin \& Konenkov 1997)
and of neutron stars accreting in  binary systems (Konar \& Bhattacharya 1997,
Urpin, Geppert \& Konenkov 1998, Urpin,  Konenkov \& Geppert 1998) the
evolution and effect of a core MF have to be investigated  too. The existence
of a core MF may explain transient features like the post--glitch  behaviour
(Alpar, Langer \& Sauls 1984, Chau, Cheng \& Ding, 1992) as well as the
presence  of residual surface fields in old neutron stars (e.g. Jahan Miri \&
Bhattacharya 1994).

It is generally accepted that a phase transition into the
super--fluid/super--conductive state takes place in the  core rather early in a
neutron star's life if the temperature drops below $10^{10}$ K (see Alpar
1991, Page 1998).  Given the typical neutron star MF strength, the core behaves as a type
II superconductor (Baym, Pethick \& Pines, 1969), i.e.  the MF penetrates the 
core in quantized flux tubes (fluxoids). The evolution of the core field is
connected with movements of the fluxoids caused by forces acting  upon
them. Under the action of these forces
the fluxoids  move towards the core--crust boundary, the magnetic flux is
expelled from the core into the crust, where the magnetic field
suffers Ohmic decay.

Ding, Cheng \& Chau 1993 (hereafter DCC) described these forces and
developed a method to calculate the radial velocity of the fluxoids, which
in turn determines the rate of flux expulsion from the core.
There are the buoyancy force (Muslimov \& Tsygan 1985), the drag force (Harvey,
Ruderman \& Shaham, 1986) and the force exerted by the neutron vortices which act
upon the fluxoids. The interaction between vortices and fluxoids is determined
by their pinning energy; in the course of neutron star spin--down the vortices
move outward and fluxoids can move faster, comove or move slower than vortices.

However, the motion of the proton flux tubes in the core leads to a distortion
of the field structure in the crust near the crust--core boundary. This
changes the  magnetic energy in the crust which, in turn, is the
source of a force that influences the flux expulsion. The
aim of this paper is to consider that ``back reaction'' of the crust onto the
fluxoid movement.

The paper is organized as follows. In section 2 we describe the forces acting
on the fluxoids in the core, the  procedure of self--consistent determination of 
the inner boundary condition for the crustal field decay, and  how to find the 
surface and core MF as functions of time. Section 3 
represents the numerical results and in section 4 we discuss the 
consequences of crustal effects for the long term evolution of isolated neutron 
stars and compare our results with those of other authors.

\section{Description of the model}

As shown by Alpar, Langer \& Sauls (1984) (see also DCC)  both the proton
fluxoids and the  neutron vortices are associated with large magnetic fields of comparable strengths.
Interaction of both structures involves a pinning energy of $E_p
\sim 10$ MeV per intersection, and the strongest  pinning occurs in the core 
regions where the flow of vortices is nearly perpendicular to the  fluxoids.

The flow velocity of the vortices, $v_n$, is governed
by the spin--down torque, which in a steady state is assumed to be determined  by 
the magnetic dipole braking of the neutron star (see DCC):

$$
v_n(t) = \frac{rk(t)\Omega_s^2(t)}{2} \;\;, \eqno(1)
$$

\noindent where

$$
k(t) = \frac{8B_e^2(t)R^6\sin^2\chi}{3Ic^3}\;\;. \eqno(2)
$$

\noindent Here, $\Omega_s$ is the angular velocity of the core
super--fluid. Its difference to the observed rotation rate of the crust,
$\Omega_c$, causes the driving Magnus force; $R$ is the radius of the neutron
star, $\chi$ is the angle between the rotational and the magnetic axis of the neutron
star, $I$ its moment of inertia and $B_e$ is the surface magnetic field in the
equatorial plane. We will assume that equation (1) holds for all distances from the
rotation axis $r \le R_c$, where $R_c$ is the radius of the super--conducting
core.

With respect to the relation between the radial velocity of the proton fluxoids, $v_p$,
(for justifications to neglect their azimuthal motion see discussion in DCC) and $v_n$
one can distinguish three different regimes. Either the vortices cross the
fluxoids ({\it forward creeping}), or both types of tubes move with the
same velocity ({\it comoving}), or the fluxoids move even faster than the
vortices ({\it reverse creeping}). Thus, in case of forward creeping the
vortices exert a force onto the fluxoids which drives them outward, while
in case of reverse creeping a resistive force acts upon the fluxoids which
counteracts the flux expulsion.  It is clear that this vortex acting force
$f_n$ depends on the angular velocity lag $\omega = \Omega_s - \Omega_c$ and
changes its direction, when the Magnus force changes its sign because the
core super--fluid rotates slower than the crust and the charged
components of the core coupled with it. The maximum lag that can be sustained by the pinning
force, $\omega_{cr} \propto E_p$, defines also the maximum force a vortex can
exert onto a fluxoid.  This force
per unit length of the fluxoid is given by (DCC, Chau, Cheng \& Ding, 1993)

$$ f_n = \frac{2 \Phi_0 \rho r
\Omega_s (t) \omega (t)}{B_c(t)} \;\;, \eqno(3) 
$$

\noindent where $\Phi_0$ denotes the quantized flux per fluxoid, $\Phi_0 =
hc/2e= 2\cdot 10^{-7}$ G$\cdot$cm$^2$, $\rho$ is the density of the matter
and $B_c$ is the mean strength of the core MF, proportional to the number of
fluxoids in the core $N_{p}$ and given by $B_c= \Phi_0 N_{p}/\pi R_c^2$.

Independently of the rotational evolution of the neutron star the fluxoids
are affected by a buoyancy force $f_b$, which is caused by the magnetic stress at the surface of the fluxoid. This force per unit length of the fluxoids is given by
(Muslimov \& Tsygan, 1985):

$$
f_b = \left(\frac{\Phi_0}{4\pi \lambda}\right)^2
        \frac{1}{R_c}\ln\left (\frac{\lambda}{\xi}\right) \;\;. \eqno(4)
$$

\noindent The relation between the London penetration length $\lambda$ and the
coherence length $\xi$ defines the condition for the formation of a type II 
superconductor, it requires that $\xi < \lambda/\sqrt{2}$.  

The drag force is caused by the scattering of the degenerate ultra--relativistic
electrons at the fluxoid MF. In an approximation which neglects collective
effects of the fluxoids, the drag force per unit length is proportional to
$v_p$ and is given by (Harvey, Ruderman \& Shaham, 1986):

$$
f_v = -\frac{3\pi}{64}\frac{n_e e^2 \Phi_0^2}{E_F \lambda}\frac{v_p}{c} \;\;, 
                                        \eqno(5)
$$

\noindent where $n_e$ is the number density of the electrons in the core, about
5\% of the neutron number density, and $E_F$ is the Fermi energy of the
electrons. For the Fermi energy as well as for $\lambda$ and $\xi$ we take the
values determined by the density at the crust--core interface, $\rho_c$. Collective
effects which modify the drag force become important only for $B_c \ge
10^{15}$ G, when the mean field becomes comparable to the field inside the
fluxoids.

\noindent Additionally, DCC take into account tension forces which result in a
coefficient of $f_b$ of the order of unity in almost the entire star. 
For simplification we will assume that the fluxoids are not bended but parallel
to each other while moving together with their roots. According to DCC, the
velocity of the fluxoids is derived from $f_n+f_b+f_v(v_p)=0$.

\noindent However, the MF which is concentrated in flux tubes in the core,
penetrates the crust too. The movement of the roots of the fluxoids leads to 
a  bending of the magnetic field lines in the crust (see figure 1). 
The power done by the core forces is equal to the net Poynting flux
through the core surface, which, in turn, is equal to the rate of change of the
sum of mechanical and field energy in the volume outside the core. We will
assume the crust to be crystallized so that the change of mechanical energy
corresponds only to the Joule heating of the crust. The contribution of the
electric field to the field energy can certainly be neglected. Thus, we find
for the balance of powers:

$$
\sum_{\rm fluxoids}\int (f_n+f_b+f_v) v_p \, {\rm d}l = 
\int\limits_{V_{\rm crust}}\frac{j^2}{\sigma}\,{\rm d}V + 
            \frac{\rm d}{{\rm d}t}\int\limits_V \frac{B^2}{8\pi}\,{\rm d}V\;. 
                             \eqno(6)
$$

\noindent Note, that $f_v$ depends on $v_p$ as well as the r.h.s. of equation\ (6). 
In the l.h.s. the integration is performed over a fluxoid's length, the summation
runs over the number of fluxoids. For sake of simplicity we consider all terms of the l.h.s. of
equation\ (6) to be position independent, take them at $r=R_c$ and approximate that
side by $(f_n+f_b+f_v) N_{p} \left< l \right> v_p$. Since the length of a
fluxoid is $l=2\sqrt{R_c^2 - r^2}$, its mean value is $\left< l \right>=
4R_c/3$. The number of fluxoids decreases with core field decay. The
multiplication with the fluxoid velocity $v_p$ yields the core  forces power which has
to be balanced by Joule heating and the change of the magnetic energy. While the
Joule heating is restricted to the volume of the crust, the change of the
magnetic energy has to be calculated both for the crust and the vacuum
environment of the neutron star. We introduce the core forces  

$$ F_{n,b,v}=f_{n,b,v}\cdot 4R_c/3 \cdot N_{p}\;\;
\eqno(7) 
$$

\noindent and the crustal force

$$
F_{crust}=- \frac {1}{v_p} 
      \left( 
        \int\limits_{V_{\rm crust}}\frac{j^2}{\sigma}\,{\rm d}V + 
            \frac{\rm d}{{\rm d}t}\int\limits_V \frac{B^2}{8\pi}\,{\rm d}V
      \right). \eqno(8)
$$

\noindent Thus, equation (6) can be rewritten in the form 
$F_n+F_b+F_v+F_{crust}=0$.

\noindent In this paper we consider the evolution of a purely poloidal dipolar
MF. Hence, the appropriate representation for the MF is that of the Stokes
stream function $S(r,t)$ which is related to the vector potential $\vec{A} =
(0, \,0, \,A_{\varphi})$ with $A_{\varphi} = S(r,t) \sin \theta /r$,
where $r$ and $\theta$ are the spherical radius and polar angle, respectively.
Then, we can express the spherical field components, $B_{r}$ and $B_{\theta}$,
in terms of $S(r,t)$, $$ B_{r} = \frac{2 S}{r^2} \cos \theta \, ,
\,\,\, B_{\theta} = - \frac{\sin \theta}{r} \cdot \frac{\partial S}{\partial
r}. \eqno(9) $$

\noindent We normalize $S(r,t)$ by its initial surface value at the equator,
$S(R,0)=B_{e0} R^2$, $s(r,t)=S(r,t)/S(R,0)$, so that $B_e(t)=B_{e0} \cdot
s(R,t)$.

\noindent While in the core there is no ohmic decay but motion of the fluxoids,
ohmic diffusion determines the field evolution in the solid crust. Thus, the field evolution in the core ($r<R_c$) is governed by

$$
\frac{\partial s}{\partial t} = -v_p \, \frac{\partial s}{\partial r}, 
                             \eqno(10)
$$

\noindent but in the crust ($R_c<r<R$) by 

$$
\frac{\partial s}{\partial t} = \frac{c^2}{4\pi\sigma}\left(\frac{\partial^2 s}{\partial r^2} 
                    - \frac{2s}{r^2}\right). \eqno(11)
$$

\noindent  Therefore, we find for the r.h.s. of equation (6) the following expressions:

$$
\int\limits_{V_{\rm crust}}\frac{j^2}{\sigma} \, {\rm d}V = 
\frac{c^2 B_{e0}^2 R^4} {6\pi} 
         \int\limits_{R_c}^R\frac{1}{\sigma}\left(\frac{\partial^2 s}
              {\partial r^2} - \frac{2s}{r^2}\right)^2 \, {\rm d}r, \eqno(12)
$$

$$
\frac{\rm d}{{\rm d}t}\int\limits_{V_{\rm crust}}\frac{B^2}{8\pi} \, {\rm d}V=
B_{e0}^2 R^4 \, \frac{\rm d}{{\rm d}t} 
 \left(\frac{2}{3}\int\limits_{R_c}^R\frac{s^2}{r^2} \, {\rm d}r +
 \frac{1}{3}\int\limits_{R_c}^R\left(\frac{\partial s}{\partial r}\right)^2 
                  {\rm d}r\right), \eqno(13)
$$

$$
\frac{\rm d}{{\rm d}t}\int\limits_{V_{\rm vacuum}}\frac{B^2}{8\pi} \, {\rm d}V =
\frac{2 B_{e0}^2 R^3} {3} \; s(R,t) \; \frac{{\rm d} s(R,t)}{{\rm d}t}. 
\eqno(14)
$$

\noindent The assumption of a homogeneous mean MF in the core leads to the 
following expression for the stream function there

$$
s(r,t) = \frac{B_c(t)}{2}\; r^2. \eqno(15)
$$

\noindent The electric conductivity in the solid crust consists mainly of
contributions from electron--phonon and electron--impurity scattering.
Electron--phonon interactions dominate the transport at high temperatures and
relatively low densities, whereas the impurity concentration determines the
conductivity at lower temperatures and larger densities. We use the numerical
data for the phonon conductivity obtained by Itoh, Hayashi \& Koyama (1993) and
an analytical expression for the impurity conductivity derived by Yakovlev \&
Urpin (1980). The crustal temperature which influences the electron--phonon
conductivity is taken from cooling curves calculated for different neutron
star models by Van Riper (1991). For the chemical composition we adopt that of
cold catalyzed matter. Note that the longterm crustal field decay in isolated neutron
stars is determined by the impurity concentration mainly which is characterized by the
impurity parameter $Q$ because after $\approx 10^6$ years the neutron star cools
down completely.  Moreover, the currents generated by flux expulsion from the
core in the crust are located in its deep high density regions. Thus, this field
evolution is almost insensitive to the cooling history of the neutron star.

\noindent The assumed homogeneity of the core MF leads to following ansatz for
the fluxoid velocity:

$$
v_p = \alpha (t) r. \eqno(16)
$$  

\noindent Thus, the solution of equation (10) is 

$$
s(r,t) = s(r,0)\exp{(-2\alpha(t)t)}, \eqno(17)
$$

\noindent where $\alpha(t)$ has to be determined by the self--consistent solution
of equation\ (6).  In that way, $v_p$ determines the inner boundary condition for
equation\ (11).

\noindent Since the MF in the center of the neutron star has to be regular we
demand as the inner boundary condition that $s(r,t)/r^2$ remains finite with $r
\rightarrow 0$, which is fulfilled by our choice of $s(r,t)$ in the core. The
boundary condition at the neutron star surface has to ensure a matching of the
interior field with the external dipolar field. That is expressed by the
requirement $R\frac{\partial s}{\partial r} = -s$ at $r=R$.

\noindent When solving equation\ (11) the boundary condition at the crust--core
boundary is given by equation\ (17) with $s(R_c,t)$, which decreases
generally non--exponentially because $\alpha$ depends on time.

\noindent In order to solve equation\ (6) we follow the ideas of DCC, however, we
have to consider additionally the crustal effects given by the r.h.s. of the
equation\ (6). By the choice of the strength of the core MF, a profile of $s(r,0)$ in the
crust, an initial rotational period and a certain EOS which determines mass,
radius and moment of inertia of the neutron star, we define the values of the
parameters at $t=0$.  

\noindent Then we start with $v_p=v_n$ or $\alpha(t)=k(t)\Omega^2/2$ where
$k(t)$ is given by equation\ (2). Adopting that equality we calculate $s(R_c,t)$ 
from equation (17) which represents the inner boundary condition for equation (11). By use of equations
(12), (13) and (14), we calculate the r.h.s. of equation\ (6), or $F_{crust}$. After
calculating $F_b$ and $F_v$, finally, we obtain $F_n$ and thus $\omega$. The core
MF defines also  the maximum lag $\omega_{cr}$ that can be sustained by the
pinning force (see equation (11) in DCC), which gives an upper limit of $F_n$. If $
-\omega_{cr} < \omega < \omega_{cr}$ then the fluxoids and vortices comove and
$v_p$, found by the procedure described above, is the real velocity of the
fluxoids.

\noindent If $\omega > \omega_{cr}$ then the fluxoids are in the forward
creeping regime.  Since $\omega_{cr}$ defines the maximum force which can be
exerted onto the fluxoid we set $\omega = \omega_{cr}$ and calculate
$F_n(\omega_{cr})$. The solution of equation\ (6) with $F_n(\omega_{cr})$ provides
$v_p$. At the forward creeping stage $v_p < v_n$, thus $v_p$ will be found as a
root of equation (6) in the interval $[0,v_n]$ using the bisectional method. If 
$\omega < -\omega_{cr}$ then the reverse creeping mode is reached. In this case we
set $\omega = -\omega_{cr}$, calculate $F_n(-\omega_{cr})$ and solve again 
equation\ (6).  This yields $v_p > v_n$.

\noindent After having determined the velocity of the fluxoids in that way we use
equations\ (11) and (17) to calculate the evolution of $s(r,t)$ and to obtain the
temporal behaviour of the MF at the crust--core boundary $B_c$ and at the
neutron star surface $B_e$. From the latter we deduce the evolution of the
spin period $P(t)=2 \pi /\Omega$. The described set of computations is performed at each time step.
 
%-------------------------------------------------------------------
\section{Numerical results}

Although the induction equation is linear, the problem is a
nonlinear one in terms of the magnetic field. Additionally, equation\ (6) which
determines the flux expulsion velocity is nonlinear in $v_p$ too. Thus, we
have to consider the time evolution for different sets of initial values of $B_e$ and
$B_c$ and to investigate their temporal evolution.

For the calculations below we use the standard cooling scenario considered by Van Riper
1991 for the neutron star model based on the Friedman--Pandharipande (FP) equation of
state (Friedman \& Pandharipande 1981) with the total mass $M=1.4M_\odot$, radius
of the star $R=10.61$ km and radius of the core $R_c=9.67$ km.  

\noindent In the present paper we consider the case that the homogeneous core
MF has initially the same strength as the surface MF (see curve 1 in figures\ 1,
2). At first we assume for both $B_e$ and $B_c$ the standard values $10^{12}$ G
and for the initial spin period $10$ ms. As described above, the expulsion of
the core MF causes distortions of the crustal field in the vicinity of the
core--crust boundary (see figures\ 1, 2), generating currents just in that
region.  Hence, the decay of the crustal field, characterized by the decay time of the 
surface MF $\tau_s$, is almost completely determined by the impurity parameter $Q$. The larger
$Q$ the more rapidly the crustal field may decay which, in turn, makes the
braking of neutron star rotation by magneto--dipole radiation less efficient. If the MF of the neutron star is confined in the crust only $\tau_s \sim 10^6 \, Q^{-1}$ years is the
appropriate estimate for the chosen neutron star model (Urpin \& Konenkov 1997). 
However, the
existence of a core MF and its expulsion into the crust causes a nonlinear
dependence of the inner boundary condition for equation\ (11) on the MF evolution and
deviations from the above estimate for $\tau_s$ will occur.

\noindent In figure\ 3 we show from top to bottom the evolution of the MF,
of the velocities of both types of flux tubes, of the forces acting upon the fluxoids, and 
of the rotational period of the neutron star, calculated by the procedure
described in the section above. Calculations were performed for three values of
the impurity parameter $Q=1$ (left column), $Q=0.1$ (middle column) and
$Q=0.01$ (right column).

\noindent Almost independently on $Q$, the velocity of the fluxoids at times
$t<10^{4.5}$ is determined by the balance of the vortex acting force $F_n$ and the
drag force $F_v$. Since the spin--down of the neutron star is very effective, the
vortices move outward fastly thereby cutting through the fluxoids: they are in
the forward creeping regime. At this stage $v_p$ falls down from $10^{-7}$ to
about $10^{-8}$ cm/s. However, the characteristic time of expulsion $\tau_e =
R_c/v_p \approx 10^6$ years, and the core MF remains almost constant. 
Nevertheless, even during that relatively short period a strong meridional field component 
$B_{\theta}$ is
generated in the crust close to the crust--core interface. This gives raise of
the crustal force $F_{crust}$, and later $v_p$ is determined by the balance of crustal and
buoyancy forces. The power--like decay of $F_n$ simply reflects the increase of
the spin period of the neutron star $P \propto \sqrt{t}$ with an almost constant
surface MF and the subsequent decrease of the vortex number density in
the core. At this stage, $v_p$ is dependent on $Q$: the lower $Q$ the lower
$v_p$, an effect of the crustal force, counteracting the expulsion.  As long as the
fluxoids are in the forward creeping regime, $B_c$ remains nearly constant. 
The forward creeping regime is followed by the comoving regime during which
$F_n$ changes its sign, and vortices and fluxoids move with the same velocity. 
Until that moment, the flux expulsion rate is governed by the balance of
crustal and buoyant forces, however, the crustal force becomes ineffective on
larger time--scales, $t >\tau_e$. This is the reason for the increase of $v_p$
and, hence, the decrease of $B_c$. Note, that the evolution of  $B_e$ follows closely that of $B_c$. The expulsion lasts till another force can prevent
it. It appears to be a common situation that at late evolutionary stages the
velocity of the fluxoids is determined by the balance of buoyancy and vortex
acting forces. The amount of flux which has to be expelled from the core in
order to reach such a balance is easy to predict by comparison of $F_n$ and
$F_b$ just before the beginning of the expulsion. Since $F_b \propto B_c$ but
$F_n \propto \sqrt{B_c}$, the residual MF strength can be estimated by

$$
\frac{B_{res}}{B_{e0}}=\left(\frac{F_n(\tau_e)}{F_b(\tau_e)}\right)^2\;\;. 
\eqno(18)
$$

\noindent The longer the time--scale of expulsion (due to the resistive effect of
the crustal force), the lower is the residual field strength. This is because
the number of vortices and $F_n$ are lower if the neutron star spins down to a
greater rotational period during a longer time. Thus, $F_n$ will be able to
balance $F_b$ on a lower MF level. In turn, the time--scale of expulsion
depends on the impurity concentration. Namely, for $Q=1,~0.1,~0.01$ the
residual field strength is about $10^{10},~10^9,~10^8$ G, while $\tau_e$
is about $10^7,~10^8,~10^9$ years, respectively. The velocity of the
fluxoids increases just after the beginning of the reverse creeping stage, and
decreases sharply when $F_b$ balances $F_n$. Concluding, we want to point out
that the main force which expels the fluxoids from the core is the  buoyancy
force. The vortex acting force influences the evolution of the MF of very young
and very old
neutron stars, while the effect of the drag force is restricted to the very
early stage only.

\noindent In figure\ 4 we show the same quantities as in figure 3 but the initial
strength of  the core and surface MF is assumed to be only $10^{11}$ G. This
leads to a drastically less  efficient spin--down of the neutron star. Because the
magnetic energy stored in the crust and in the surrounding vacuum is  $\propto B_e^2$, 
the influence of
the crustal forces on the expulsion is much weaker than in the case of
$B_{e0}=10^{12}$ G. For a comparatively low field strength the situation is
similar  to that considered by DCC because practically only $F_b$, $F_n$ and
$F_v$ determine the  field evolution. Due to the much slower spin--down, from
the beginning both vortices and  fluxoids are moved with the same velocity.
Again, in the young neutron star ($t<10^6$ years),  $F_n$ and $F_v$ are much
greater than $F_{crust}$ and $F_b$, while in the old neutron star $F_n$ and 
$F_b$ are the dominant forces. Since the crustal force is much weaker, $\tau_e
<< \tau_s$,  i.e. the evolution of $B_c$ is decoupled from that of $B_e$.

\noindent The evolution of core and surface MF and of the spin period for a
strong initial  MF, $B_{e0}=10^{13}$ G, is shown in figure 5. In this case the
strong crustal force prevents the  flux expulsion from the core much more
effectively. The strong and long living MF causes an  efficient spin--down, up
to $30 - 300$s, depending on the impurity concentration. The residual  field
strength is about $3\cdot 10^7$ G for $Q=1$, and about $3\cdot 10^6$ G for
$Q=0.1$.  For even lower impurity concentrations $\tau_e$ exceeds the Hubble
time. Note that the large initial MF enforces $\tau_e=\tau_s$. This is a consequence 
of the strong crustal force, which decelerates effectively the expulsion and couples the evolution of the core with that surface MF.

\noindent Generally, the residual field strength is anti--correlated with the
initial field  strength and positively correlated with $Q$. Once the balance of
crustal and buoyant forces is replaced by the balance of vortex
acting and buoyant forces on a much lower level, both $B_c$ and $B_e$ decay
down to a residual value, determined by the final rotational period $P_f$ of
the neutron star. For such neutron stars which do not reach their $P_f$ even after
$10^{10}$ years (as for  $B_{e0}=10^{13}$ G, $Q=0.01$ and
$B_{e0}=10^{11}$ G, $Q=1$) a residual MF is not attained.

\noindent Among the processes considered here are two dissipative ones: the
work done by the  drag force and the Joule heating produced by field decay in
the crust. Assuming that all heat  produced in that way is irradiated
from the surface one can estimate the  corresponding surface temperature $T_s$
by use of the relation

$$
\dot{Q}=4\pi R^2 \sigma_{SB} T_s^4\;\;, \eqno(19)
$$

\noindent where  $\sigma_{SB}$ is the Stephan--Boltzmann constant,
$\dot{Q}=F_v\cdot v_P+\dot{Q}_{Joule}$, and
$\dot{Q}_{Joule}$ is given by equation\ (12) (see Miralles et al. 1998 for the
discussion of the validity of this equation).  In figure\ 6 the temporal evolution
of $T_s$ is presented for different initial MF strengths and the values of
the impurity parameter considered above. When the standard cooling scenario for
a FP--neutron star applies it becomes clear that the contribution of those
dissipative processes is considerable during the photon cooling era in
relatively old ($t> 10^6$ years) radio--pulsars.

\section{Discussion}

We considered the effect of the neutron star crust onto the expulsion of a core
MF and its ohmic decay in the crust. To this aim we solved self--consistently the
equation of balance of the powers of forces acting on the fluxoids in the core and the 
rate of change of magnetic energy outside the core of the neutron star,
assuming a homogeneous MF in the core for all the life of the neutron star.

\noindent The evolution of $B_e$ determines the rotational evolution of the
neutron star, which in turn has considerable effects on the vortex acting
force defining the residual field strength.

\noindent It turns out that the characteristic time--scale of the decay of the surface MF
$\tau_s$ increases with increasing initial MF strength and decreasing impurity
parameter. The amount of $B_e$--decay at $t \ge \tau_s$ is correlated with the
spin--down rate. A slow spin--down to $P_f \approx 1$ s results in a small field
decay by less than two orders of magnitude ($B_{e0}=10^{11}$ G, $Q=0.01$) while
a drastic spin--down to $P_f \approx 300$ s yields a field decay by about
seven orders of magnitude ($B_{e0}=10^{13}$ G, $Q=0.1$).

\noindent Comparing our results with those obtained for a purely crustal field
decay in isolated neutron stars (Urpin \& Konenkov 1997), we find a
considerable deceleration of the decay of a field penetrating the entire star. 
The field evolution for the FP model with standard cooling and $Q=0.01$ in the
case of crustal MF yields an impurity dominated decay with $\tau_s \approx 10^8$ years which becomes then nearly exponential. 
In the present model, only for the small
initial field of $10^{11}$ G, $\tau_s \approx 10^8$ years, while it is $10^9$
years for  $B_{e0}=10^{12}$ G and is in the order of the Hubble time scale for
$B_{e0}=10^{13}$ G. Evidently, for a purely crustal MF with the inner boundary
condition $s(R_c,t)=0$ valid for all neutron star's life, no residual field can
be obtained. The existence of a residual field as well as the strong dependency
of $\tau_s$ on $B_{e0}$ reflects the nonlinear mutual dependency of the field
and rotational evolution, governed by the balance of forces acting upon the
fluxoids.

\noindent Bhattacharya \& Datta, 1996, studied the decay of a neutron star MF
just expelled from the core and deposited in the bottom layers of the crust.
They found a rather strong decrease of the final MF strength after $10^{10}$
years with increasing impurity parameter. This result can not be confirmed by
our investigation: {\it a larger $Q$ results in a higher residual field}. This is due
to the fact that for a larger $Q$ the crustal force is less strong, the  time--scale
of expulsion is shorter, the surface field decays faster, the  spin--down is less
efficient and  the residual field determined by balance of the $F_n$ and $F_b$ is
stronger.

\noindent Taking into account that under the assumption of a purely crustal MF a value of $Q=0.1 ... 0.01$ is at least not
in disagreement with observations (Urpin \& Konenkov 1997),
these values taken for our model result in a constant $B_e$ for almost all
conceivable pulsar lifetimes; even for $B_{e0}=10^{11}$ G and $Q=1$ a remarkable
field decay would start only for $t > 10^6$ years. Note that the early ($t \le 10^6$ years) cooling--determined crustal field evolution is included in our
investigations. Only extremely large impurity parameters (perhaps $Q > 1$,
which would reflect qualitative deviations from the bcc crystalline structure
of the crust) would allow for a field decay in $t < 10^6$ years. Thus assuming
an initial field strength larger than $10^{11}$ G, our study yields a $\tau_s >
10^7$ years, in agreement with  the statistical results for isolated
radio--pulsars found by Bhattacharya et al. (1992) and Hartman et al. (1996).

\noindent It is clearly seen from figure\ 3, that the main force which is
responsible for expulsion of the flux from the core is the  buoyancy force. The
hypothesis about the so called "spin--down induced" expulsion of magnetic flux
(Konar, Bhattacharya 1998 and references therein) seems to be adequate only in
case of a weak ($\sim 10^{11}$ G) initial magnetic field. If the magnetic field is
stronger, say, $10^{12}$ G, vortices will cut through the fluxoids, whereas the 
latter are anchored in the crust for the time of $\ge 10^7$ years even in case
of high impurity concentration ($Q=1$). During this time the neutron star spins
down by a factor of about 100, while $B_c$ remains almost the same, in
contradiction with the core field evolution predicted by the spin--down induced
mechanism of expulsion. 

\noindent In the present paper we study the magneto--rotational evolution
of isolated neutron stars suffering a spin--down by magneto--dipole
radiation. We found that for relatively large field strengths, $10^{12} {\rm G} \le B_{e0} \le 10^{13} {\rm G}$,
the residual field $B_{\rm res} \propto P_f^{-2}$, in accordance
with the result of DCC. We also found that in this range of $B_{e0}$ the expulsion times--scale is determined mainly by the balance of buoyant and crustal
forces. These forces are independent of the spin--down rate of the neutron
star, i.e. the expulsion time--scale does not depend on the specific
braking mechanism. Jahan Miri \& Bhattacharya, 1994, studied the evolution of neutron stars in binaries. They adopted the hypothesis on the spin--down induced magnetic flux expulsion and found $B_{\rm res} \propto P_{max}^{-1}$, where $P_{max}$ is the maximum period reached by the neutron star during the propeller phase. We do not expect that our model (if applied to the neutron star evolution in binaries) will confirm this result.  

\noindent 
In the context of the current discussion of magnetars, which are thought to be 
highly magnetized isolated neutron stars spinning down by magneto--dipole braking, 
the investigation of the crustal effect onto the expulsion of initial MFs $> 10^{14}$G 
is an urgent task. Our present model is not suitable for that purpose since we consider 
only ohmic diffusion in the crust. In the case of extremely large field strength as 
expected for magnetars, the field  evolution in the crust is more complicated, effectively 
accelerated by a Hall cascade (Goldreich \& Reisenegger 1992) and/or fracturing of the 
crust (Thompson \& Duncan 1996). Thus, in order to describe the flux expulsion in magnetars our 
model has to be modified qualitatively.

\noindent
Both the investigation of flux expulsion in magnetars and the effects of accretion on the 
evolution of a MF permeating the whole neutron star, will be considered in forthcoming papers.

\section*{Acknowledgment}

The work of D.K. was supported in part by the Russian Foundation of Basic Research under
the Grant 97-02-18096(a) and by INTAS under the grant 96-0154.  D.K. gratefully
acknowledges the kind hospitality of  the Nordic Institute for Theoretical Physics
(Nordita) and of Prof. Raedler's group of Astrophysikalisches Institut Potsdam.  It is
also pleasure to thank Prof. C. Pethick for valuable and stimulating discussions.

\section*{References}

\noindent
Alpar, M.A. 1991,
in Neutron Stars: Theory and Observation,
ed Ventura, J. \& Pines D.
(Dordrecht: Kluwer Academic Publishers) 49

\noindent
Alpar, M.A., Langer S.A., Sauls J.A. 1984
ApJ 282,533

\noindent
Baym, G., Pethick, C., Pines, D. 1969, Nature, 224, 673

\noindent
Bhattacharya, D., Datta B. 1996
M.N.R.A.S. 282, 1059 

\noindent
Bhattacharya, D., Wijers, R.A.M.J., Hartman, J. W., Verbunt, F. 1992
A\&A 254, 1992

\noindent
Blandford R., Applegate J., Hernquist L., 1983, MNRAS, 204, 1025

\noindent
Chau, H. F., Cheng, K. S., Ding, K. Y. 1992
ApJ 399, 213

\noindent
Ding, K. Y., Cheng, K. S., Chau, H. F. 1993
ApJ 408, 167

\noindent
Friedman, B., Pandharipande, V. 1981, Nucl. Phys. A, 361, 502

\noindent
Geppert, U., Page, D., Zannias, T. 1999,
A\&A 345, 847

\noindent
Geppert, U., Urpin, V. 1994
M.N.R.A.S. 273, 490

\noindent
Goldreich, P., \& Reisenegger, A. 1992, ApJ, 395, 250

\noindent
Haberl, F., Motch, C., Buckley, D.A.H., Zickgraf, F.-J., Pietsch, W. 1997
A\&A 326, 662 

\noindent
Hartman, J., Bhattacharya, D., Wijers, R., Verbunt, F. 1996, A\&A, 322, 477

\noindent
Harvey, J., Ruderman, M., Shaham, J. 1986,
Phys. Rev. D. 33, 2084

\noindent
Itoh, N., Hayashi, H., Koyama, Y. 1993,
ApJ 418, 405

\noindent
Jahan Miri, M., Bhattacharya, D. 1994,
M.N.R.A.S. 269, 455

\noindent
Konar, S., Bhattacharya, D. 1997,
M.N.R.A.S. 284, 311

\noindent
Konar, S., Bhattacharya, D. 1998, 
astro-ph/9812035

\noindent
Muslimov, A., Tsygan, A. 1985,
Ap\&SS 115, 41

\noindent
Page, D. 1998,
in Neutron Stars and Pulsars,
eds N. Shibazaki, N. Kawai, S. Shibata, \& T. Kifune, 
(Universal Academy Press: Tokyo), p. 183.

\noindent
Thompson, Ch., Duncan, R. 1993,
ApJ, 408, 194

\noindent
Thompson, Ch., Duncan, R. 1996,
ApJ, 473, 322

\noindent
Urpin, V., Konenkov, D. 1997,
M.N.R.A.S. 292, 167

\noindent
Urpin, V., Konenkov, D., Geppert, U. 1998,
M.N.R.A.S. 299, 73

\noindent
Urpin, V., Geppert, U., Konenkov, D. 1998,
M.N.R.A.S. 295, 907

\noindent
Urpin, V., Levshakov, S., Yakovlev, D. 1986,
M.N.R.A.S. 219, 703

\noindent
Van Riper, K. 1991,
ApJS 75, 449

\noindent
Wiebicke, H.-J., Geppert, U. 1996,
A\&A 309, 203

\noindent
Yakovlev, D., Urpin, V. 1980,
Sov. Astron. 24, 303

\newpage
\section*{Figure captions}

\noindent {\bf Fig.\ 1} Due to the motion of fluxoids from position 1 to
position 2 the magnetic field lines are bended in the crust. This causes a
change of the magnetic energy and heat release within the crust. Thus, a force
raises which can counteract the movement of fluxoids.

\noindent {\bf Fig.\ 2} The same as in figure 1 but in terms of the normalized
stream function $s(r,t)$. The gradient of $s$ near the crust--core interface
corresponds, according to equation\ (9), to a generation of a $\theta$--component 
of the crustal magnetic field.

\noindent {\bf Fig.\ 3} The temporal evolution of magnetic field, forces,
velocities of both kinds of tubes and spin period of the neutron star for
$B_{e0}=10^{12}$ G. Left, middle and right columns correspond to the different
values of the impurity parameter: $Q=1, 0.1, 0.01$, respectively.

\noindent {\bf Fig.\ 4} The same as in figure 3 but for $B_{e0}=10^{11}$ G.

\noindent {\bf Fig.\ 5} The temporal evolution of the magnetic field and spin
period for $B_{e0}=10^{13}$ G. The numbers at the curves correspond to the 
different values of the impurity parameter $Q$.

\noindent {\bf Fig.\ 6} The temporal evolution of the surface temperature for
different values of $B_{e0}$ and $Q$. Curve 1: no additional heating, curve
2: $B_{e0}=10^{11}$ G, $Q=0.1$; curve 3: $B_{e0}=10^{11}$ G, $Q=0.01$; curve 4:
$B_{e0}=10^{12}$ G, $Q=0.01$; curve 5: $B_{e0}=10^{12}$ G, $Q=0.1$; curve 6:
$B_{e0}=10^{13}$ G, $Q=1$.

\newpage
\begin{figure}[t]
\vspace{9cm}
\input epsf
\epsfbox {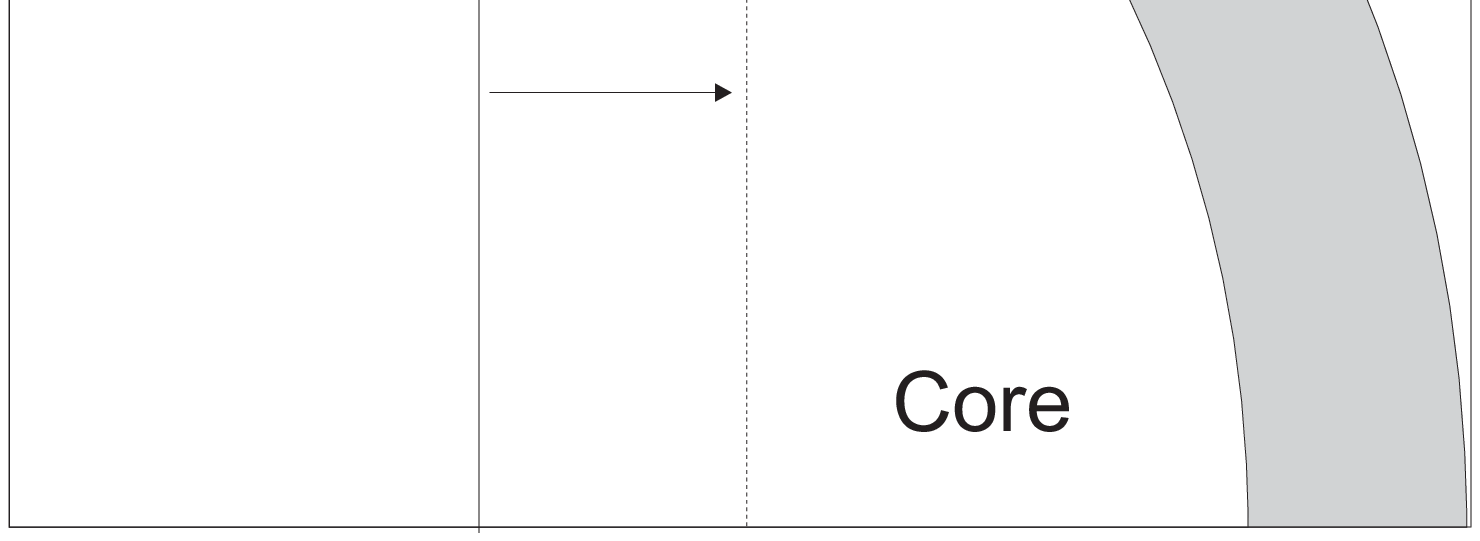}

\caption{ Due to the motion of fluxoids from position 1 to
position 2 the magnetic field lines are bended in the crust. This causes a
change of the magnetic energy and heat release within the crust. Thus, a force
raises which can counteract the movement of fluxoids.}

\end{figure}

\newpage
\begin{figure}[t]
\vspace{9cm}
\input epsf
\epsfbox {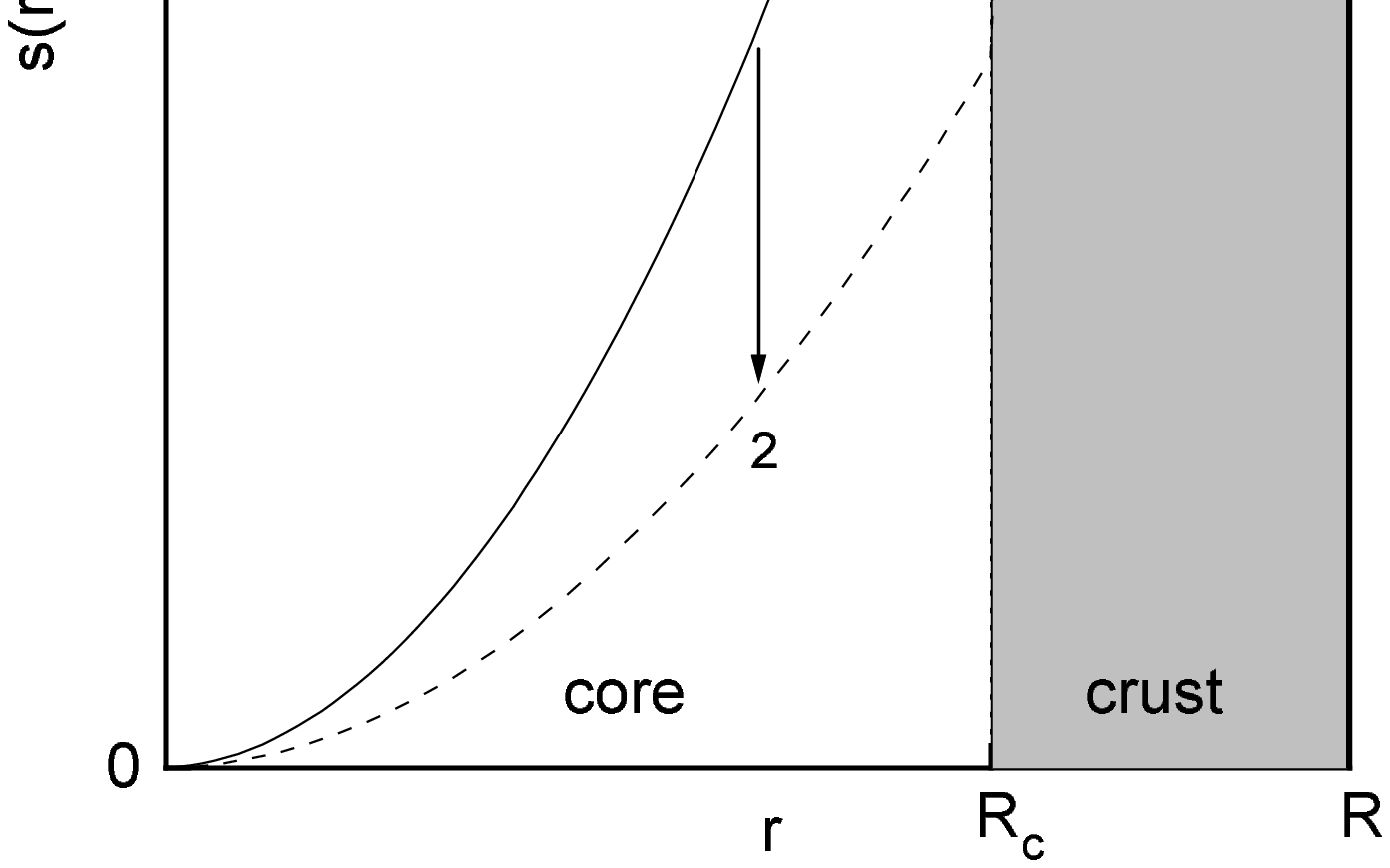}

\caption{ The same as in figure 1 but in terms of the normalized stream function
$s(r,t)$. The gradient of $s$ near the crust--core interface corresponds to a
generation of a $\theta$--component of the crustal magnetic field.}

\end{figure}

\newpage
\begin{figure}[t]
\vspace{12cm}
\input epsf
\epsfbox {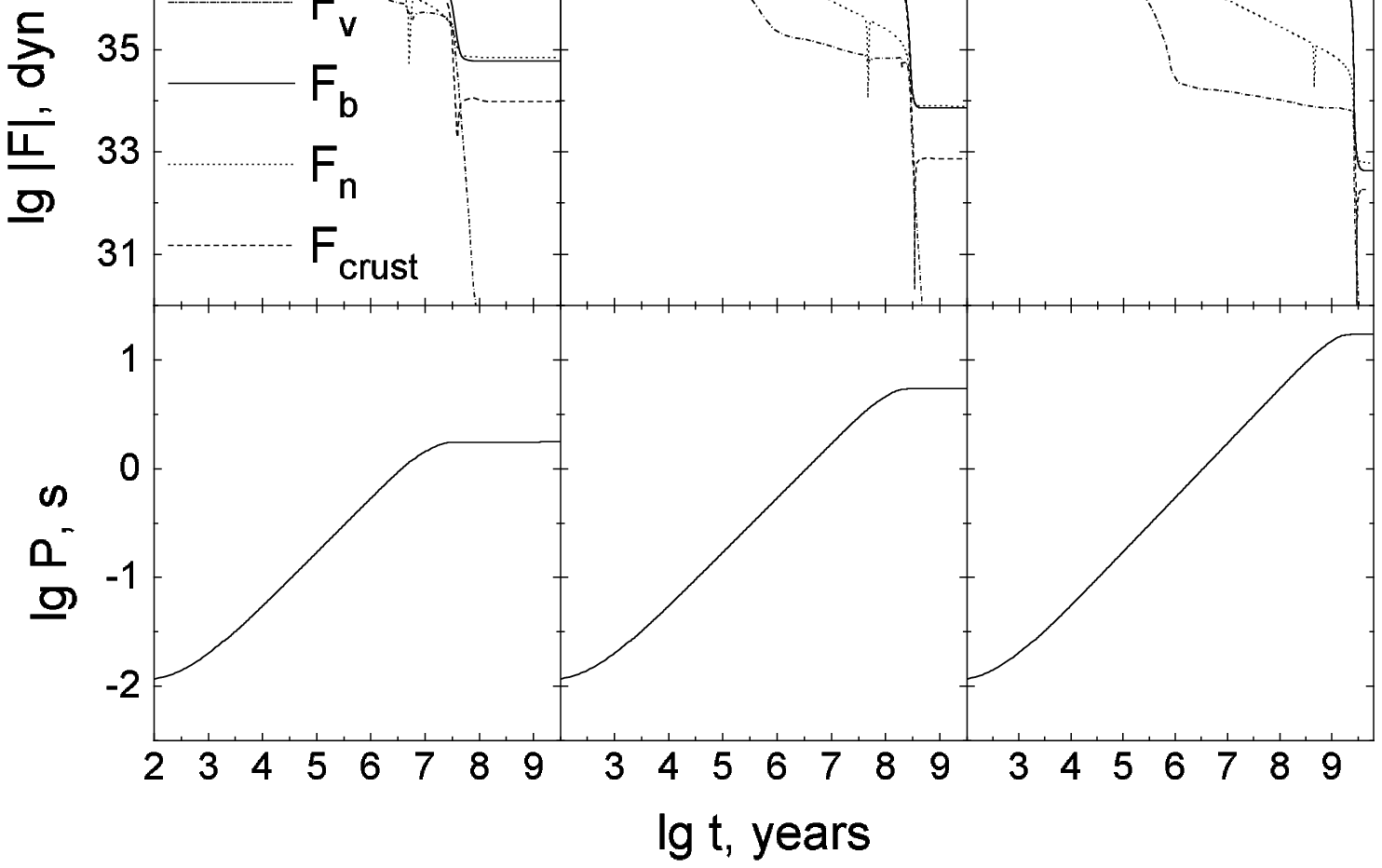}

\caption{ The temporal evolution of magnetic field, forces, velocities of
both kinds of tubes and spin period of the neutron star for
$B_{e0}=10^{12}$ G. Left, middle and right columns correspond to the different
values of the impurity parameter: $Q=1, 0.1, 0.01$, respectively.}

\end{figure}

\newpage
\begin{figure}[t]
\vspace{12cm}
\input epsf
\epsfbox {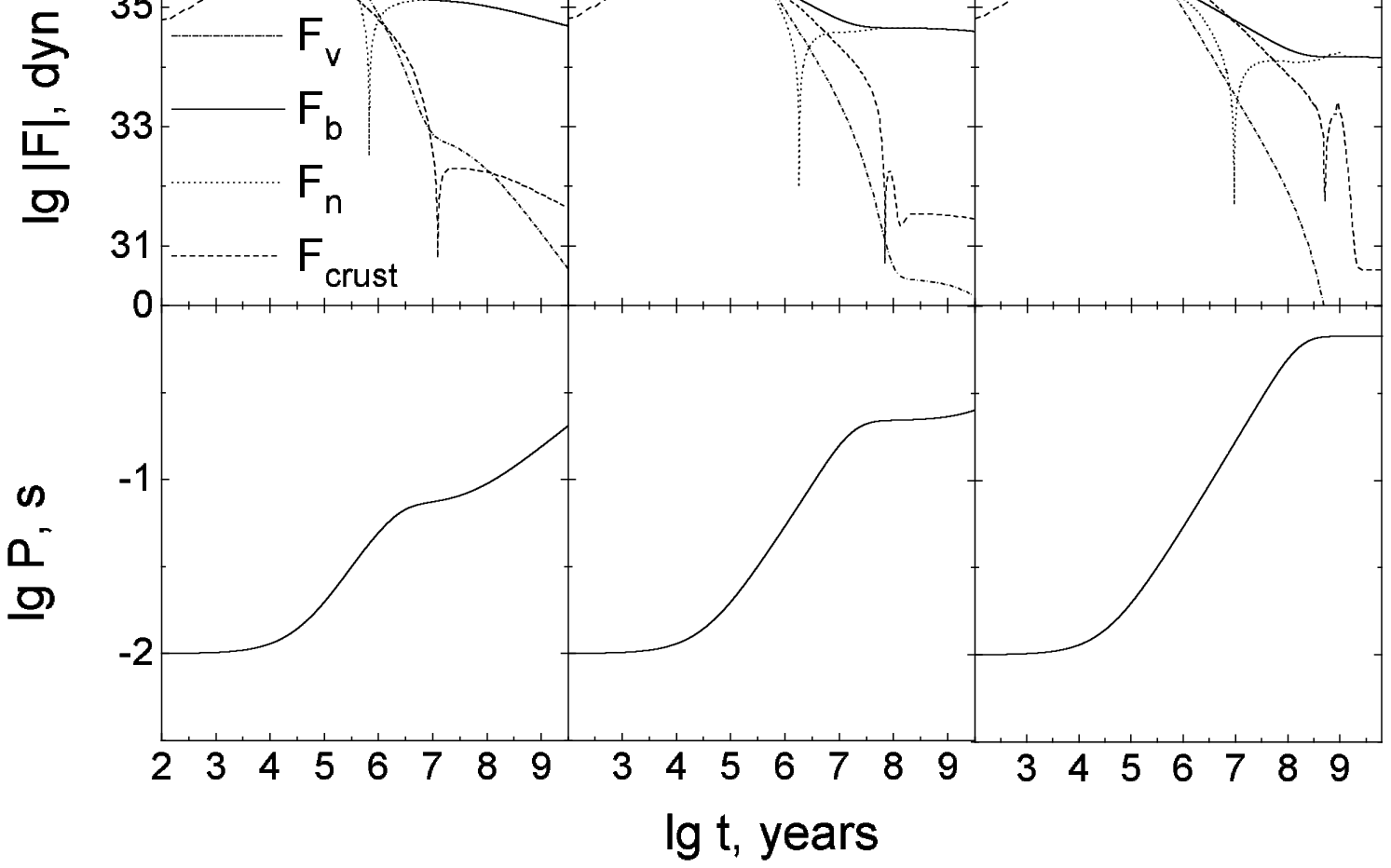}
\caption{ The same as in figure 3 but for $B_{e0}=10^{11}$ G.}
\end{figure}

\newpage
\begin{figure}[t]
\vspace{9cm}
\input epsf
\epsfbox {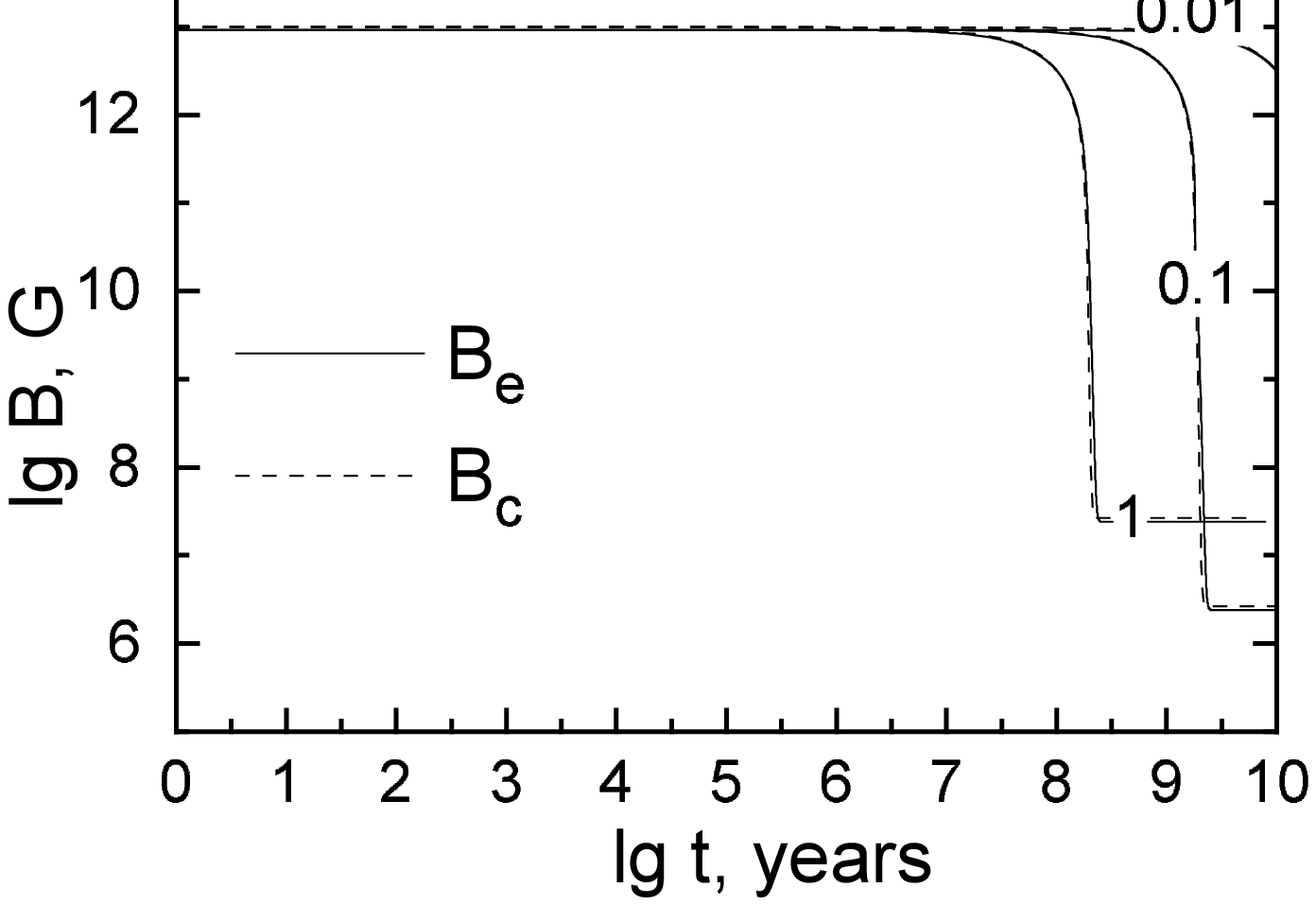}
\caption{ The temporal evolution of the magnetic field and spin period for 
$B_{e0}=10^{13}$ G
for different values of the impurity parameter.}
\end{figure}

\newpage
\begin{figure}[t]
\vspace{9cm}
\input epsf
\epsfbox {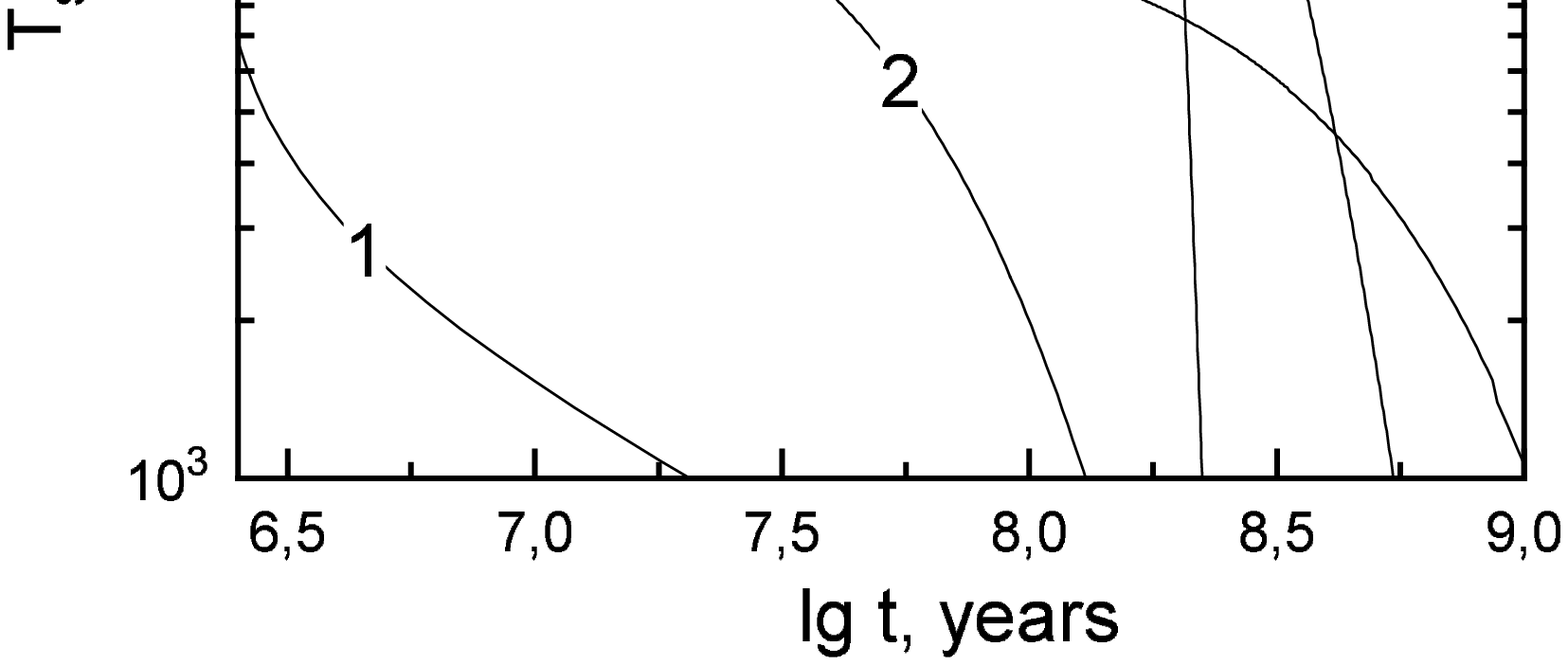}
\caption{ The temporal evolution of the surface temperature for different
values of $B_{e0}$ and $Q$. 
Curve 1: no additional heating, 
curve 2: $B_{e0}=10^{11}$ G, $Q=0.1$;
curve 3: $B_{e0}=10^{11}$ G, $Q=0.01$;
curve 4: $B_{e0}=10^{12}$ G, $Q=0.01$;
curve 5: $B_{e0}=10^{12}$ G, $Q=0.1$;
curve 6: $B_{e0}=10^{13}$ G, $Q=1$.}
\end{figure}

\end{document}